\documentclass{aa}  
\usepackage{graphicx}
\usepackage{txfonts}
\begin{document}
   \title{Studying the small scale ISM structure with supernovae}
  \subtitle{}
   \author{F. Patat\inst{1}
   \and
    N.L.J. Cox\inst{2,3} 
    \and 
   J. Parrent\inst{4,5}
    \and
    D. Branch\inst{4}
}

   \offprints{F. Patat}

   \institute{European Organization for Astronomical Research in the 
	Southern Hemisphere (ESO), K. Schwarzschild-str. 2,
              85748, Garching b. M\"unchen, Germany\\
              \email{fpatat@eso.org}
              \and
	      Herschel Science Centre, European Space Astronomy Centre, 
	      ESA, PO Box 78, 28691 Madrid, Spain 
              \and 
              Institute of Astronomy, K.U. Leuven,
              Celestijnenlaan 200D, 3001 Leuven, Belgium
              \and
              Department of Physics and Astronomy. University of Oklahoma, 
              Norman, OK 73019 
              \and
              Department of Physics and Astronomy, Dartmouth College, 
              Hanover, NH 03755-3528
}

   \date{Received 08/01/2010; accepted 25/02/2010}

\abstract{}{}{}{}{} 
% 5 {} token are mandatory
 
  \abstract              
   %{The spatial scales of the ISM on scales smaller than 100 AU are very poorly
   %known, and are mostly based on column density variations observed along 
   %the line of sight to bright Galactic stars, on time scales of many years.}  
   {} 
   {In this work we explore the possibility of using the fast expansion of     
    a Type Ia supernova photosphere to detect extra-galactic ISM column        
    density variations on spatial scales of $\sim$100 AU on     
    time scales of a few months.}                                              
   {We constructed a simple model which describes the expansion of the     
    photodisk and the effects of a patchy interstellar cloud on the observed equivalent  
    width of \ion{Na}{i} D lines. Using this model we derived the behavior       
    of the equivalent width as a function of time, spatial scale and amplitude 
    of the column density fluctuations.}                                       
   {The calculations show that isolated, small ($\leq$100 AU) clouds with 
    \ion{Na}{i} column densities exceeding a few 10$^{11}$ cm$^{-2}$ 
    would be easily detected. In contrast, the effects of a more realistic, 
    patchy ISM become measurable in a fraction of cases, and for peak-to-peak 
    variations larger than $\sim$10$^{12}$ cm$^{-2}$ on a scale of 1000 AU.} 
   {The proposed technique provides a unique way to probe the extra-galactic 
    small scale structure, which is out of reach for any of the methods used 
    so far. The same tool can also be applied to study the sub-AU Galactic ISM 
    structure.}


\authorrunning{F. Patat et al.}
\titlerunning{Studying the small scale ISM structure with supernovae}

   \maketitle
%
%________________________________________________________________

\section{\label{sec:intro}Introduction}

For many years it was accepted that the minimum size for the column
density fluctuations in the Galactic ISM is around 1 pc
($\sim$2$\times$10$^5$ AU). The common understanding was that although
sub-parsec structures do exist, only a tiny fraction of the column
density could be ascribed to these small scales (Dickey \& Lockman
\cite{dickey}). However, the pioneering VLBI work by Dieter, Welch \&
Romney (\cite{dieter}), and the later confirmation by Diamond et
al. (\cite{diamond}) demonstrated the existence of significant
fluctuations over scales of 20 AU. These findings were confirmed by 21
cm absorption measurements against high-velocity pulsars (Frail et
al. \cite{frail91,frail94}), which showed that the \ion{H}{i} column
density varies significantly over scales between 5 and 110 AU, with
10\%-15\% of the cold neutral gas distributed in AU-sized structures
(Frail et al. \cite{frail94}). However, some more recent radio
observations on the same pulsars (Weisberg \& Stanimirovi\'c
\cite{weisberg}) have shown that the variations are far smaller than
those originally found by Frail et al. (\cite{frail91,frail94}).

These studies were followed by a series of works looking at the
variations of \ion{Ca}{ii} and/or \ion{Na}{i} column densities along
the lines of sights to close binaries or high proper motion stars (see
Crawford \cite{crawford03} and Lauroesch \cite{lauroesch07} for a
review).  Similar investigations were carried out for molecular gas
(CH, CH$^+$, and CN; Pan et al. \cite{pan}; Rollinde et
al. \cite{rollinde}), and diffuse interstellar bands (Cordiner, Sarre
\& Fossey \cite{cordiner}).  An alternative method is the study of
interstellar absorptions along the lines of sight to stellar clusters,
like M92 (Andrews, Meyer \& Lauroesch \cite{andrews}) and $\omega$-Cen
(Van Loon et al. \cite{vanloon}), or the Magellanic Clouds
(e.g. Andr\'e et al. \cite{andre}). For a general review on the small
ISM structures in our Galaxy the reader is referred to Haverkorn \&
Goss (\cite{sins}).

In this article we present an independent technique to analyze
extra-galactic ISM structure on spatial scales of about 100 AU.  The
proposed method is based on the extremely high expansion velocity
displayed by a supernova (SN) photosphere ($\sim$10$^4$ km s$^{-1}$ or
5.7 AU day$^{-1}$). A Type Ia SN reaches a photospheric radius of
$\sim$10$^{15}$ cm ($\sim$100 AU) in two weeks from the explosion, and
expands at a rate of 6 to 3 AU day$^{-1}$ during the first two months
of its evolution.  If the typical size of the fluctuations in an
intervening cloud is much larger than 10$^{15}$ cm, then the
associated absorption features will not evolve with time. On the
contrary, if the ISM is patchy on comparable scales, the
column density fluctuations will translate into measurable variations
of the corresponding absorption features.

After introducing a simple model for the calculation of time-dependent
line equivalent widths for a given cloud geometry
(Sect.~\ref{sec:model}), we present the results of Monte-Carlo
simulations (Sect.~\ref{sec:results}), and discuss the applicability
of the method and the effects on the observations of Type Ia's
(Sect.~\ref{sec:disc}). Appendix A gives the details on the derivation
of the composite equivalent width.

\section{\label{sec:model}A simple model}

\begin{figure}
\centerline{
\includegraphics[width=7cm]{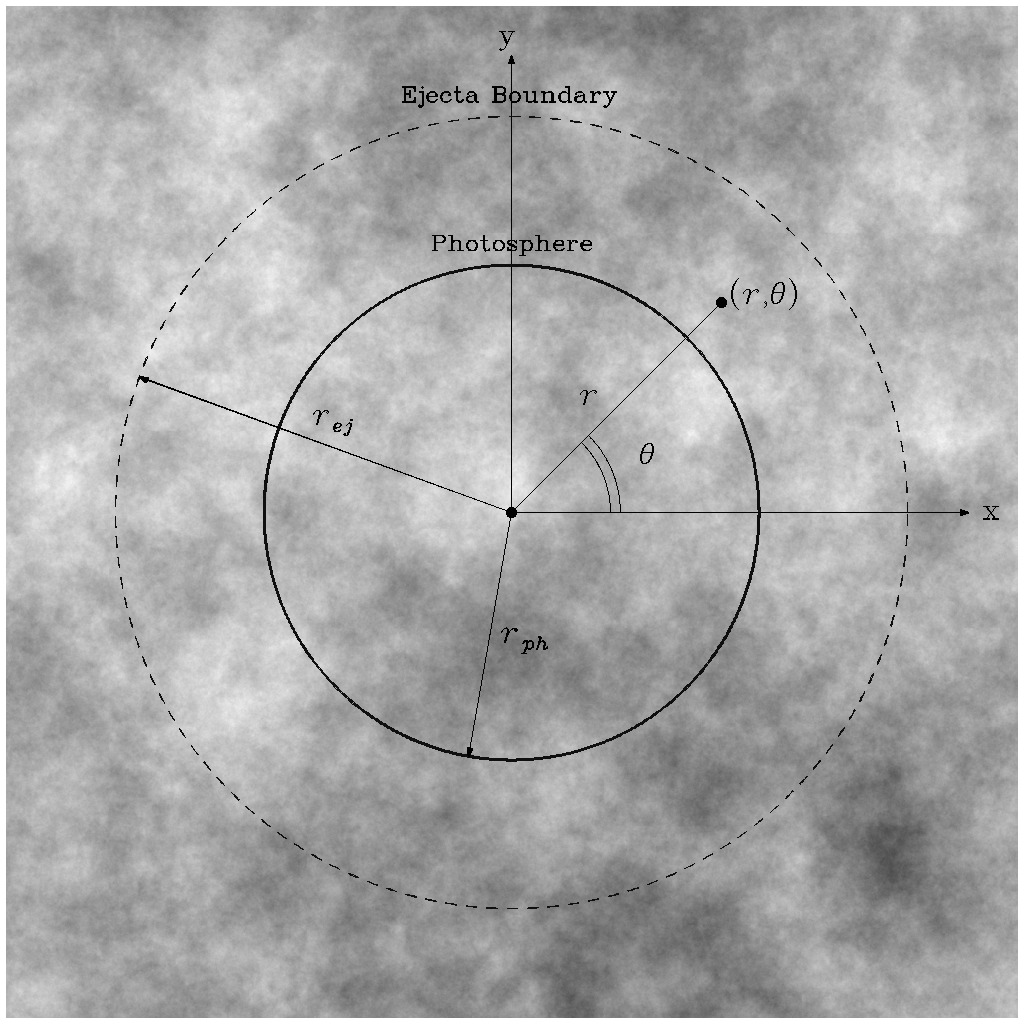}
}
\caption{\label{fig:cloud}Relevant quantities used in the text. The
  underlying column density map was generated using a power-law
  spectrum with $\gamma$=$-$2.8 (see Sect.~\ref{sec:cloud}).}
\end{figure}

Let us fix a reference polar coordinates system ($r$,$\theta$) whose
origin is located at the explosion center (see
Fig.~\ref{fig:cloud}). As seen from a far observer (the distance
between the SN and the observer is assumed to be much larger than that
between the SN and the intervening material), the SN will appear as an
expanding photodisk which extends to the ejecta boundary radius
$r_{\rm ej}(t)$. We then consider a cloud placed in front of the SN,
at a distance large enough that the explosion has no effect on its
physical conditions (i.e. $>$10 pc, see Simon et
al. \cite{simon09}. See also Sect.~\ref{sec:disc} here), and we
indicate with $N(r,\theta)$ the cloud column density in the species
under consideration. Finally, we introduce $\Phi(r,t)$ as the
time-dependent surface brightness profile of the photodisk at the
wavelength of interest\footnote{We assume $\Phi$ has no azimuthal
  dependence, i.e. that the SN is spherically symmetric. For a Type Ia
  SN this is a reasonable assumption (Wang \& Wheeler
  \cite{araa}).}. This function is normalized as follows

\begin{displaymath}
\int_o^{r_{\rm ej(t)}} \int_0^{2\pi} \Phi(r,t)\; \;r \; d\theta \; dr = 1\; ,
\end{displaymath}

\noindent so that the total continuum flux emitted by the photodisk
along the line of sight is equal to unity. If $g(N,b)$ is the curve of
growth for the given transition and Doppler parameter $b$, the
equivalent width ($EW$) produced by an infinitesimal cloud column with
cross section $dA=r\;d\theta\;dr$ is $dEW = g[N,b] \;
\Phi(r,\theta)\;dA$.

If we neglect the small contribution by photons scattered by the cloud
into the line of sight, the total equivalent width is then computed
integrating the contribution of each single infinitesimal element over
the photodisk (see Appendix A for the details):

\begin{displaymath}
EW(t)=\int_o^{r_{\rm ej(t)}} \int_0^{2\pi}
g[N(r,\theta),b(r,\theta)]\; \Phi(r,t)\; \;r \; d\theta \; dr\;.
\end{displaymath}

So far we have considered the possibility that the Doppler parameter
$b$ can change across the cloud. However, in the lack of evidence for
significant variations over the scales of interest (see for instance
Welty \& Fitzpatrick \cite{welty}), in the following we will assume
that $b$ is constant across the relevant portion of the cloud. We will
briefly discuss the effects of a space-dependent Doppler parameter in
Sect.~\ref{sec:results}.

With the aid of the outlined procedure, one can follow the time
evolution of $EW$ for any input cloud column density map, provided the
photodisk's expansion law is known.

In the assumption of homologous expansion (see for instance Jeffery \&
Branch \cite{jeffery}), the radius of the photosphere $r_{\rm ph}(t)$
is obtained multiplying the photospheric velocity $v_{\rm ph}(t)$ by
the time $t$ elapsed from the explosion. For $v_{\rm ph}(t)$ we
adopted a best fit to values computed via SYNOW modeling of the
spectroscopically normal SN~1994D, a standard Type Ia event (Branch et
al. \cite{branch05}). During the photospheric phase ($t<$100 days),
$v_{\rm ph}(t)$ is well approximated by an exponential law, and the
best fit relation takes the following form: 

\begin{displaymath}
r_{\rm ph}(t) = \left (3.5 + 5.3 \; e^{-t/36.5}\right) \; t, 
\end{displaymath}

\noindent where $r_{\rm ph}$ is expressed in AU and $t$ in days from
the explosion. This law shows that the change in the photodisk
dimensions is significant during the first two months: from day $-$10
to day +50 (counted from maximum light) $r_{\rm ph}$ increases by more
than a factor of 5 (60 AU to 320 AU). 

\begin{figure}
\centerline{
\includegraphics[width=8cm]{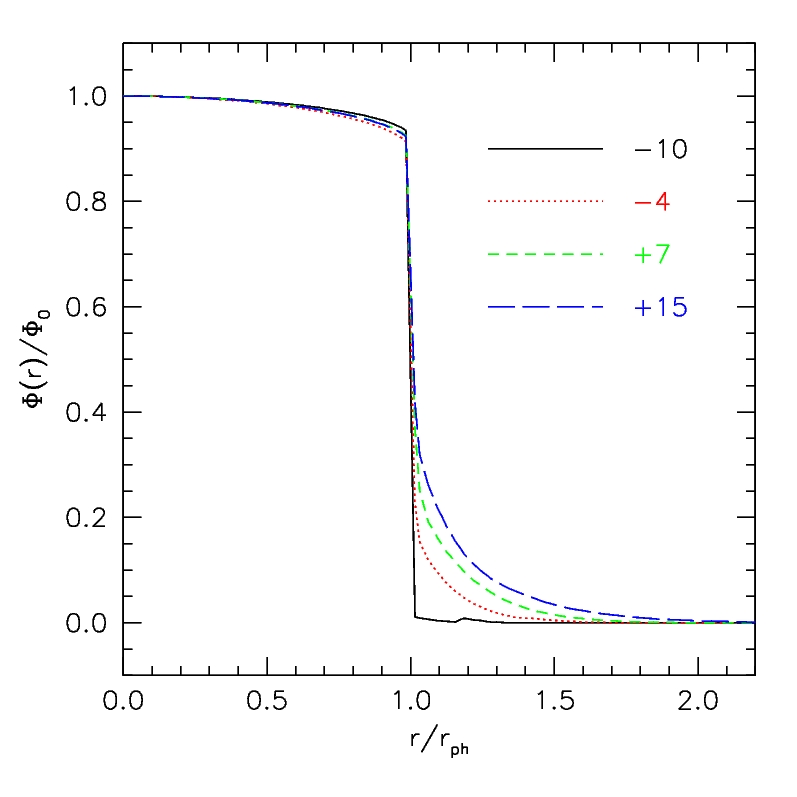}
}
\caption{\label{fig:prof}Surface brightness profiles derived from
  SYNOW best fits of SN~1994D spectra on four different epochs ($-$10,
  $-$4, +7 and +15 days from $B$ maximum).}
\end{figure}

As for the photodisk surface brightness profile, this was computed
using a modified version of the spectrum synthesis code
SYNOW\footnote{Note that SYNOW, like some other SN spectrum synthesis
  codes (see e.g. Mazzali \& Lucy \cite{mazzali}), has zero limb
  darkening.}  (Branch et al. \cite{branch05}).  The profile, obtained
from best fits of SN~1994D spectra for the \ion{Na}{i} D rest-frame
wavelength, rapidly drops at $r=r_{\rm ph}$ during the early phases
(see Fig.~\ref{fig:prof}). As time goes by, a significant fraction of
the flux (up to $\sim$16\% on day +15) is emitted above the
photosphere, and it is due to scattering by the broad \ion{Na}{i} D
doublet intrinsic to the SN (see for instance Jeffery \& Branch
\cite{jeffery}). At all epochs, $\Phi$=0 for $r/r_{\rm ph}\geq$2.6,
which we used as the effective external boundary of the ejecta. To
include the time dependence we tabulated $\Phi(r)$ for a number of
epochs ($-$10, $-$4, +7, +15, +28 and +50 days from $B$ maximum) and
subsequently used a linear interpolation to derive the profile at any
given epoch. The time from $B$ maximum light was converted into $t$
using the rise time of SN~1994D (18 days; Vacca \& Leibundgut
\cite{vacca}). In view of the lack of very early spectra, we
conservatively assumed that $\Phi$=0 for $r>r_{\rm ph}$ at t=0.

\subsection{\label{sec:cloud}Cloud generation}

We explored two possible cases: {\it i)} an isolated, homogeneous,
spherical cloud with radius $r_C$ and offset $x_C$ with respect to the
line of sight; {\it ii)} a patchy sheet with an input spectrum for the
column density fluctuations.

In the first case, the column density profile is $N(r^\prime)=N_0
\sqrt{1-(r^\prime/r_C)^2}$, where $r^\prime$ is the projected distance
from the cloud center ( $r^\prime\leq r_C$), and $N_0$ is the column
density corresponding to a ray going through the center of the
cloud. The average column density is $\langle N\rangle$=2/3 $N_0$. As
the case of an isolated, small cloud is probably quite unrealistic, it
may be rather regarded as a simplified model of an over-density on an
otherwise homogeneous sheet.

For the patchy sheet we adopted a procedure similar to the one
described by Deshpande (\cite{deshpande00b}). Molecular clouds
(Elmegreen \& Falgarone \cite{elmegreen}), the diffuse ionized
component (Cordes et al. \cite{cordes}), and \ion{H}{i} (Stanimirovi\'c et al.
\cite{stanimirovic}; Deshpande, Dwarakanth \& Goss
\cite{deshpande00a}) have a fractal structure, characterized by a
power-law behavior. Therefore, if $k=l_0/l$ is the wave-number
corresponding to a given spatial scale $l$ (where $l_0$ is the maximum
spatial scale under consideration), the power spectrum of the
fluctuations can be written as $P(k)\propto k^\gamma$. For the large
scale structure of \ion{H}{i} in the SMC $\gamma\simeq-$3 (Stanimirovi\'c et
al. \cite{stanimirovic}) and $\gamma$=$-$2.75 for the cold atomic gas
in the Galaxy (Deshpande et al. \cite{deshpande00a}). Since the
variations observed on scales of $\sim$100 AU and below can be
explained in terms of a single power spectrum description (Deshpande
\cite{deshpande00b}), we computed the column density maps using a
power-law with $\gamma$=$-$2.8.  After generating a power spectrum map
$P(k_x,k_y)$ in the Fourier plane (with $k=(k^2_x +k^2_y)^{1/2}$), we
derive the real and imaginary parts of the Fourier transform of
$N(x,y)$ using $\sqrt{P(k)}$ as the modulus of the complex numbers,
while phases are generated as random numbers uniformly distributed
between 0 and 2$\pi$. The column density map is then obtained
anti-transforming into ordinary space. Given the radius of the
photosphere during the time interval typically covered by observations
($<$400 AU), we adopt $l_0$=1024 AU.

\begin{table}
\tabcolsep 2.7mm
\caption{\label{tab:err} Estimated Equivalent Width RMS errors
  $\sigma_{EW}$ computed using Equation~\ref{eq:sigew} 
({\it SNR}=100 and $N$=10$^{12}$ cm$^{-2}$).}  
\centerline{
\begin{tabular}{ccccccc}
\hline
      &          & \multicolumn{5}{c}{$\delta \lambda$ (\AA\/ pix$^{-1}$)}\\
\cline{3-7}
$FWHM$       & $b$          & 0.005& 0.01& 0.02& 0.05& 0.10\\
(km s$^{-1}$)& (km s$^{-1}$) &      &     &     &     &     \\
\hline
     & 1.0   &  0.28  &  0.42  &  0.63  &  1.18  &  1.03 \\
3.0  & 3.0   &  0.35  &  0.50  &  0.74  &  1.28  &  1.83 \\
     & 5.0   &  0.45  &  0.65  &  0.95  &  1.69  &  2.42 \\
\cline{2-7}
     & 1.0   &  0.37  &  0.52  &  0.77  &  1.43  &  1.86 \\
5.0  & 3.0   &  0.40  &  0.57  &  0.81  &  1.53  &  2.51 \\
     & 5.0   &  0.48  &  0.69  &  1.01  &  1.69  &  2.42 \\
\cline{2-7}
     & 1.0   &  0.45  &  0.63  &  0.89  &  1.67  &  2.53 \\
7.0  & 3.0   &  0.46  &  0.66  &  0.93  &  1.75  &  2.51 \\
     & 5.0   &  0.52  &  0.75  &  1.06  &  1.90  &  2.99 \\
\cline{2-7}
     & 1.0   &  0.52  &  0.74  &  1.07  &  1.88  &  2.53 \\
9.0  & 3.0   &  0.53  &  0.76  &  1.10  &  1.75  &  3.07 \\
     & 5.0   &  0.58  &  0.83  &  1.17  &  1.90  &  2.99 \\
\hline
 \multicolumn{7}{l}{Note: errors are expressed in m\AA.}
\end{tabular}
}
\end{table}

Once the two-dimensional column density map is generated, it is
re-normalized to have an average column density $\langle N\rangle$ and
peak-to-peak fluctuations $\Delta N$ (with $\Delta N\leq 2\langle
N\rangle$). An example cloud is presented in Fig.~\ref{fig:cloud}. If
$\Delta N(l)$ is the column density difference between two points
separated by a distance $l$, the numerical simulations show that for
$\gamma$=$-$2.8 the RMS value of $\Delta N(l_0)$ is $\sim$10\% of the
maximum variation on the same spatial scale. Furthermore, the
differences at higher wave-numbers decrease proportionally to
$k^{(\gamma+2)/2}$, as predicted by theory (Deshpande
\cite{deshpande00b}). This implies that the fluctuations expected on
scales of 100 AU are $\approx$40\% of those observed on scales of 1000
AU, which for \ion{Na}{i} can reach a few 10$^{12}$ cm$^{-2}$ (Andrews
et al. \cite{andrews}). Therefore, a significant amount of structure
is expected on spatial scales comparable to the typical photospheric
radius.

\subsection{Measurement error estimates}

The ability of detecting small $EW$ variations hinges on the precision
to which $EW$s can be measured. In turn, this relates to the
instrumental setup and the signal-to-noise ratio {\it SNR} reached on
the adjacent continuum per resolution element. To estimate the
  expected RMS uncertainty $\sigma_{EW}$ we have used the following
  formula (Chalabaev \& Maillard \cite{chalabaev}; see their
  Equation~A.10), derived for photon noise-dominated conditions and with
  no assumption on the line profile:

\begin{equation}
\label{eq:sigew}
\sigma^2_{\rm EW}=N \; \frac{\delta^2 \lambda}{SNR^2}\; \frac{F}{F_c}\;
+ \frac{\sigma^2_{F_c}}{F_c^2} \left ( \Delta \lambda - EW\right)^2
\end{equation}

\noindent where $\Delta \lambda$ is the line integration range,
$\delta \lambda$ is the dispersion (\AA\/ px$^{-1}$), $N=\Delta
\lambda/\delta \lambda$ is the number of pixels corresponding to the
integration range, $F$ is the average flux within the absorption,
$F_c$ is the average continuum level, and $\sigma_{F_c}$ is the
associated uncertainty. In addition to the parameters directly related
to the instrumental setup, the application of this analytical
expression requires the knowledge of other quantities which depend on
the line profile. To compute them for a given column density, we
generate synthetic Voigt profiles with input velocity parameter and a
sampling equal to $\delta\lambda$. This is then convolved with a
Gaussian profile with a given full width half-maximum ($FWHM$) to
mimic the instrumental broadening. Using this profile we calculate the
line integration range $\Delta \lambda$, which we define as the
wavelength interval in which the equivalent width is 99\% of
total. Within this interval we finally derive the average line flux
$F$.  As for the continuum $F_c$, we assume this is computed on two
symmetric, equally extended regions on either side of the
absorption. If $\Delta \lambda_c$ is the total extension of this
region, then the number of pixels used to determine $F_c$ is
$N_c=\Delta\lambda_c/\delta\lambda$, so that
$\sigma^2_{F_c}/F_c^2=1/(N_c\; SNR^2)$. In the computations we have
used $\Delta \lambda_c$=2 \AA\/ which, given the featureless nature of
the SN pseudo-continuum, is a realistic value\footnote{For the
  purposes of measuring the $EW$ of an inter-stellar line, the continuum
  definition in a SN spectrum is much easier than in a stellar
  spectrum, where the presence of other intrinsic features may
  contaminate the adjacent regions.}.

The results obtained for $N$=10$^{12}$ cm$^{-2}$, and {\it SNR}=100
for different values of $FWHM$, $b$ (1, 3 and 5 km s$^{-1}$
corresponding to $EW$ of 58, 113 and 138 m\AA\/ respectively ), and
$\delta \lambda$ are presented in Table~\ref{tab:err}. As the RMS
errors are inversely proportional to {\it SNR}, these values can be
readily scaled to different signal-to-noise ratios. These results have
been checked against Monte-Carlo simulations and were found to be
consistent to within a few 0.1 m\AA. Incidentally, this questions the
need for a revision of the Chalabaev \& Maillard formula discussed by
Vollmann \& Eversberg (\cite{vollmann}).

In the following we will consider an equivalent width variation
$\Delta EW$ detectable if $|\Delta EW|\geq 5\sqrt{2} \sigma_{EW}$. For
a typical case where $FWHM$=7 km s$^{-1}$, $b$=1 km s$^{-1}$, $\delta
\lambda$=0.01 \AA\/ pix$^{-1}$, and {\it SNR}=100, this turns
  into a 5-$\sigma$ detection limit $\Delta EW_{lim}$=4.4 m\AA\/
  ($\Delta EW_{lim}$=5.3 m\AA\/ for $b$=5 km$^{-1}$).

\section{\label{sec:results}Results of simulations}

Although the model can be used for any inter-stellar absorption line,
in the following we present the results obtained for \ion{Na}{i}
D$_2$, because it is a strong transition, it falls in a region almost
free of telluric absorption features, and in a spectral interval where
most optical, high-resolution spectrographs have their maximum
sensitivity.

\begin{figure}
\centerline{
\includegraphics[width=8cm]{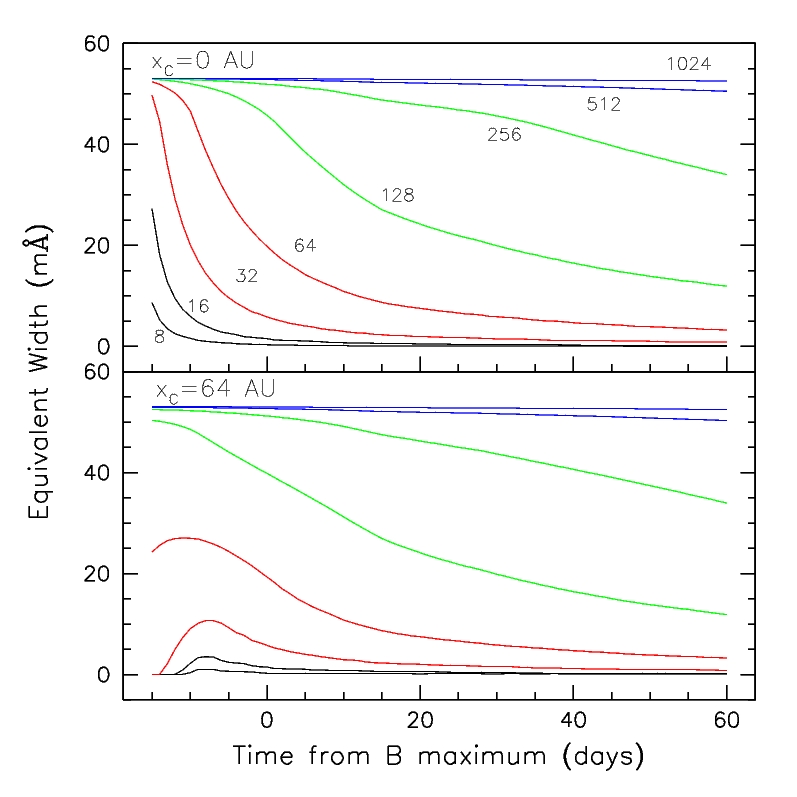}
}
\caption{\label{fig:offsphere}Examples of simulated \ion{Na}{i} D$_2$
  $EW$ variation as a function of time for a spherical, homogeneous
  cloud with offset $x_C$=0 (upper panel) and $x_C$=64 AU (lower
  panel).  The various curves refer to different cloud radii $r_C$ (8,
  16, 32, 64, 128, 256, 512 and 1024 AU from bottom to top). In all
  cases $N_0(\ion{Na}{i})$=5$\times$10$^{11}$ cm$^{-2}$ and $b$=1 km
  s$^{-1}$.}
\end{figure}

\subsection{Isolated spherical cloud}

Example $EW$ evolutions for two different cloud offsets $x_C$ (0 and 64
AU) and a number of cloud radii $r_C$ are presented in
Fig.~\ref{fig:offsphere} up to 2 months after maximum light. In
general, the maximum variability is expected when the cloud is close
to the center of the photodisk. The maximum variation is achieved for
cloud radii between 64 and 128 AU. Also, small clouds are better
detected during the early phases (when their size is comparable to
that of the photosphere), while the detection of large clouds requires
a larger time span.  If the cloud is too large ($r_C>$512 AU), then
the $EW$ variation is not sufficiently ample to be detected. With a
minimum set of two observations taken 10 days apart in the pre-maximum
phases, for typical high-resolution setup, a 5-$\sigma$ detection
limit of 4.4 m\AA, $b$=1 km s$^{-1}$ and $N_0$=5$\times$10$^{11}$
cm$^{-2}$, the simulations show that one is able to detect clouds with
$r_C$ between 16 and 128 AU up to a maximum offset of 64 AU. For
larger offsets the cloud starts to intersect the photodisk when its
size is too large and the corresponding covering factor is too
small. Besides implying probably unrealistic density contrasts,
increasing the column density does not enhance the detectability of a
small size cloud, since this rapidly becomes totally opaque. We note
that, while this causes the saturation of the covering factor, it does
not produce a saturated profile in the emerging absorption line.

A distinctive feature of small ($r_C\leq$64 AU), offset clouds is an
$EW$ growth followed by a decrease during the pre-maximum light phase
(Fig.~\ref{fig:offsphere}, lower panel). This is due to the growth of
the covering factor as the photodisk starts intersecting the
off-centered knot. Once a maximum value is reached, the subsequent
increase in the photodisk size causes the covering factor to
drop. Although this mechanism can produce an absorption feature which
grows in strength and then disappears on timescales of a month, this
is expected to happen only during the pre-maximum epochs, when the
relative increase in the photodisk surface is very fast.  Larger
clouds placed at larger offsets ($x_C\geq$256 AU) produce features
which start appearing around maximum light, but keep steadily
increasing in strength up to several months after maximum.

Finally, for central column densities (or density contrasts) smaller
than $\sim$2$\times$10$^{10}$ cm$^{-2}$, the $EW$ variations are always
below the detection limit.

\subsection{\label{sec:patchy}Patchy clouds}

Because of the stochastic geometry of the power-law clouds, their
effects were evaluated using a statistical approach. For a given cloud
realization we computed $EW(t)$ and derived the absolute peak-to-peak
variation $\Delta EW$ over the whole time interval $-$10 to +50 days.
To mimic a more realistic situation, we did this also on two
sub-sets of data-points including only post-maximum observations (0,
+10, +20, +30, +40, +50 and 0, +10, +20, +30 days). As statistical
estimators we computed the median absolute peak-to-peak variation
over the full range ($\overline{\Delta EW}$), the semi-inter-quartile
range and the 99-th percentile ($\Delta EW_{99}$). For the three time
ranges we finally estimated the 5-$\sigma$ detection
probabilities $P_t$ ($-$10$\leq t \leq$+50), $P_0$ ($t\geq$0) and
$P_{30}$ (0$\leq t \leq$+30). The results are presented in
Table~\ref{tab:sim} for different values of $\langle N \rangle$,
$\Delta N$ and $b$. For each parameter set 5000 cloud realizations
were computed (blank values indicate detection probabilities
$<$2$\times$10$^{-4}$). $\langle EW\rangle$ indicates the $EW$
corresponding to $\langle N\rangle$ and the input Doppler parameter.

Visual inspection of a number of realizations showed that the $EW(t)$
curves are smooth, with typical timescales of the order of 10
days. The variation rate is systematically larger at early epochs and
decreases as time goes by. The simulations show that peak-to-peak
column density fluctuations smaller than 10$^{11}$ cm$^{-2}$ on a
scale of $\sim$1000 AU do not produce any measurable effects. Even
with high ($\sim$100) signal-to-noise ratio observations, starting 10
days before maximum and covering the first two months of the SN
evolution, the detection probability is below 2\%. This grows
significantly when the fluctuations exceed $\sim$10$^{12}$
cm$^{-2}$. Incidentally, this implies that SNe suffering higher
extinction are expected to display more pronounced variations (see
also Chugai \cite{chugai}).  Because of the saturation effect, the $EW$
variations are more marked for larger values of the Doppler
parameter. Finally, for column densities exceeding $\sim$10$^{13}$ cm
$^{-2}$ the detection probability decreases due to line saturation.

\begin{table}
\tabcolsep 2.0mm
\caption{\label{tab:sim}Results of Monte-Carlo simulations for power-law 
clouds with $\gamma$=$-$2.8. Equivalent width variations are peak-to-peak 
(see Sect.~\ref{sec:patchy}).}
\begin{tabular}{cccccccc}
\multicolumn{8}{c}{$\gamma$=$-$2.8, $b$=1.0 km s$^{-1}$, $\Delta EW_{lim}$=4.4 m\AA}\\
\hline
$\langle N\rangle$ & $\Delta N$ & $\langle EW \rangle$ &$\overline{\Delta EW}$ & $\Delta EW_{99}$ &
  $P_t$ & $P_0$ & $P_{30}$ \\
\multicolumn{2}{c}{(10$^{12}$ cm$^{-2}$)} & m\AA & m\AA & m\AA &\%  & \%  & \% \\  
\hline
0.1 & 0.1 & 16.3 & 0.6 (0.3) & 2.2 & -    & -   & -   \\
0.1 & 0.2 &      & 1.3 (0.6) & 4.5 & 1.5  & -   & -   \\
1.0 & 1.0 & 57.7 & 0.7 (0.4) & 2.7 & -    & -   & -   \\
1.0 & 2.0 &      & 1.5 (0.8) & 8.0 & 5.0  & 2.1 & 1.1 \\
10.0&10.0 & 85.7 & 0.6 (0.3) & 1.9 & -    & -   & -   \\ 
10.0&20.0 &      & 1.1 (0.5) & 5.1 & 1.8  & 0.6 & 0.2 \\ 
\hline
\multicolumn{8}{c}{}\\
\multicolumn{8}{c}{$\gamma$=$-$2.8, $b$=5.0 km s$^{-1}$, $\Delta EW_{lim}$=5.2 m\AA}\\
\hline
0.1 & 0.1 & 18.9 & 0.9 (0.4) &  3.1 & -    & -    &  -  \\
0.1 & 0.2 &      & 1.7 (0.8) &  6.2 &  2.3 &  0.1 &  -  \\
1.0 & 1.0 & 138.0& 4.6 (2.1) & 15.6 & 40.7 & 22.2 &  9.8\\
1.0 & 2.0 &      & 9.1 (4.3) & 33.9 & 77.1 & 56.1 & 40.3\\
10.0&10.0 & 334.0& 3.1 (1.4) & 11.1 & 19.2 &  7.1 &  2.4\\ 
10.0&20.0 &      & 6.3 (3.1) & 32.6 & 60.3 & 40.6 & 25.2\\ 
\hline
\multicolumn{8}{l}{Note: values in parenthesis indicate the semi-inter-quartile range.}
\end{tabular}
\end{table}

To study the effect of a spatially variable Doppler parameter, we have
run a set of simulations in which $b$ is allowed to fluctuate around
the average value across the cloud. We have tentatively modeled the
Doppler parameter map using the same algorithm and spatial scales
spectrum adopted for the column density generation (see
Sect.~\ref{sec:cloud}).  We remark that this is not meant to reproduce
real physical conditions, but only to estimate the consequences of
velocity dispersion fluctuations (for instance, $b$ and $\langle N
\rangle$ were left completely independent). The MC runs (with 1$\leq b
\leq$5 km s$^{-1}$) show that for $\langle N\rangle<$10$^{11}$
cm$^{-2}$ there is practically no difference with respect to the case
with constant Doppler parameter. The differences start to be
significant at $\langle N\rangle>$10$^{12}$ cm$^{-2}$, for which the
typical variations and detection probabilities become much larger.

This was to be expected, since in the quasi-linear regime attained at
low column densities the effect tends to average out. On the
contrary, as one enters the non-linear part of the curve of growth,
the largest variations are produced by the regions of the cloud where
the Doppler parameter is higher (i.e. less subject to saturation),
thus skewing the distribution towards more marked $EW$ fluctuations.
The exact behavior depends on the way the velocity dispersion varies
across the cloud and how this (if any) relates to the column density
fluctuations. However, the conclusion that a variable Doppler
parameter enhances the detection probability is of general validity.

\section{\label{sec:disc}Discussion and Conclusions}

The simulations presented in this paper indicate that marked time
effects on the measured $EW$s are expected for small ($r_C\leq$100
AU), isolated clouds with $N_0(\ion{Na}{i})\sim$10$^{11}$-10$^{12}$
cm$^{-2}$ and for small offsets ($x_C<$100 AU). However, the existence
of such structures is seriously questioned in terms of pressure
equilibrium arguments and the yet unknown processes that would produce
them (see Heiles \cite{heiles} and references therein). Frail et
al. (\cite{frail94}) detected maximum $\Delta N(\ion{H}{i})$
variations that range from $\sim$10$^{19}$ to $\sim$5$\times$10$^{20}$
cm$^{-2}$ on scales between 5 and 100 AU. For a Galactic
\ion{Na}{i}/\ion{H}{i} ratio (Ferlet, Vidal-Madjar \& Gry
\cite{ferlet}) this turns into $\Delta N(\ion{Na}{i})$ between
5$\times$10$^{10}$ and 3$\times$10$^{12}$ cm$^{-2}$.  These large
changes have been interpreted as arising within ubiquitously
distributed small structures. However, this picture has been
questioned by Deshpande (\cite{deshpande00b}), who has convincingly
shown that the observations are consistent with a single power-law
description of the ISM, down to AU scales. In these circumstances,
peak-to-peak variations $\Delta N(\ion{Na}{i})\sim$10$^{12}$ cm$^{-2}$
are expected on scales of $\sim$100 AU. Our calculations (see
Table~\ref{tab:sim}) show that for a Type Ia SN observed under the
most favorable conditions ({\it SNR}$>$100 on all epochs, spanning
from $-$10 to +50 days) these would appear in less than 10\% of the
cases for $b\sim$1 km s$^{-1}$. This fraction increases to $\sim$80\%
for $b\sim$5 km s$^{-1}$, but in all cases it is $\Delta EW\leq$40
m\AA. These small variations imply negligible changes in $E(B-V)$, but
can have some effect on observing programmes studying the evolution of
\ion{Na}{i} features possibly arising in the circumstellar environment
of Type Ia progenitors (Patat et al. \cite{patat07}; Simon et
al. \cite{simon09}). However, we note that these values are more than
a factor of 10 smaller than the \ion{Na}{i} D variations detected in
the Type Ia SN~2006X (Patat et al. \cite{patat07}), which were
attributed to the ionization effects induced by the SN on its
circum-stellar environment. In contrast, no statistically significant
variations were detected for the CN, CH, CH$^+$, \ion{Ca}{i} lines and
DIBs associated to an interstellar cloud in the host galaxy (Patat et
al. \cite{patat07}; Cox \& Patat \cite{cox}). In the only other well
studied case published so far (SN~2007le), the $EW$ of four \ion{Na}{i}
D components remained constant to within a few m\AA\/ during six
epochs spanning about 3 months (Simon et al. \cite{simon09}). In this
time range $r_{\rm ph}$ changed approximately from 100 to 400 AU, and
the lack of evolution is in line with the predictions of our model for
a power spectrum ISM and definitely excludes the presence of small,
isolated clouds with sizes comparable to $r_{\rm ph}$.  Although the
available data are still scanty, the multi-epoch, high-resolution
campaigns which are being conducted for the study of Type Ia
progenitors will provide a more statistically significant sample.

All the discussion so far is based on the assumption that the physical
conditions of the ISM are not modified by the SN explosion, so that
all variations in the absorption lines are due to pure geometric
effects. Indeed, a Type Ia SN can produce changes in the ionization
balance of low-ionization species (like \ion{Na}{i} or \ion{K}{i}) up
to quite large distances, of the order of 10 pc (Patat et
al. \cite{patat07}; Chugai \cite{chugai}; Simon et
al. \cite{simon09}). Given the UV flux predicted for these distances
(Simon et al. \cite{simon09}), the ionization timescale of \ion{Na}{i}
is expected to be about 120 days. Since the electron density in the
ISM is low, the recombination time is extremely long. Therefore, under
the assumption of a constant ionizing field, one would expect the
amount of neutral Na to decrease with timescales of several months,
hence mimiking small scale structure effects.  However, the UV flux of
a Type Ia SN decreases significantly after maximum light (a factor
$\sim$20 in the first 40 days; Brown et al. \cite{brown}) implying
that the maximum distance is probably less than 10 pc. Given the range
of possible distances to an inter-stellar cloud within the host galaxy, this
suggests that significant \ion{Na}{i} column density variations in the
ISM induced by the SN radiation field are improbable.

Another important fact is that, because of the much higher ionization
potential of \ion{Ca}{ii}, and the strong UV line blocking present in
Type Ia spectra, the $EW$ of the ubiquitous H\&K lines
becomes insensitive to the SN radiation field already at distances of
a few 0.1 pc (Simon et al. \cite{simon09}). In contrast, in the case
of a geometrical origin, all species are expected to show synchronous
variations, although possibly with different amplitudes and time
scales (Lauroesch \& Meyer \cite{lauroesch03}).  Therefore, a
comparison between the behaviors of \ion{Na}{i} and \ion{Ca}{ii}
should allow one to disentangle between geometrical and ionization
effects, similar to what has been proposed by Patat et
al. (\cite{patat07}).

The method we presented enables the study of small scale structure in
the extragalactic ISM, which is out of reach for any of the techniques
deployed so far. In this respect we note that the same method can be
in principle applied to the Galactic ISM to study its sub-AU
structure, for which no direct measurements are available
yet. Although probably requiring very high signal-to-noise ratios,
this technique might put important constraints on the very small scale
structure.  In this article we have discussed the case of the strong,
easily detectable \ion{Na}{i} D lines. However, other weaker lines
(e.g., \ion{K}{i}, \ion{Ca}{i}, \ion{Ca}{ii}) can be used, especially
when the \ion{Na}{i} D lines are saturated.  In general, the
simultaneous study of different atomic/diatomic lines along the lines
of sight to SNe will contribute to get a more detailed picture of the
physical conditions of the ISM in the small scales regime.

%%%%%%%%%%%%%%%%%%%%%%%%%%%%%%%%%%%%%%%%%%%%%%%%%%%%%%%%%%%%%%%%%%%%%
%
%            A P P E N D I X
%
%%%%%%%%%%%%%%%%%%%%%%%%%%%%%%%%%%%%%%%%%%%%%%%%%%%%%%%%%%%%%%%%%%%%%
\appendix

\section{\label{sec:app}Calculation of composite equivalent width}

Let us consider an extended photodisk, a cloud placed in front of it
and an absorption line with profile function $\phi(\nu)$,
normalized so that

\begin{displaymath}
\int_0^{+\infty} \phi(\nu) \; d\nu = 1.
\end{displaymath}

If $N$ is the space-dependent column density of the cloud (atoms
cm$^{-2}$) for the atomic species under consideration, the
monochromatic line optical depth is $\tau(\nu) = K\;N\;\phi(\nu)$,
where $K$ is the frequency-integrated line opacity (cm$^{2}$
atom$^{-1}$). Let us now consider an infinitesimal element of the
photodisk $dA=r\; d\theta \; dr$ in the polar reference system
introduced in Sect.~\ref{sec:model} (see Fig.~\ref{fig:cloud}), and
define $\Phi(r)$ as the radial profile of continuum surface brightness
for the photodisk. For the sake of simplicity we assume that the
continuum $I_0$ is constant across the line profile and is normalized
to unity:

\begin{displaymath}
I_0 = 2\pi \int_0^{r_{ej}} \Phi(r)\; r\; d\theta\; dr = 1\;.
\end{displaymath}

With these settings, the monochromatic intensity contributed by
the infinitesimal photodisk element $dA$ to the total observed
intensity can be written as

\begin{displaymath}
dI(\nu,r,\theta) = e^{-K\; N(r,\theta)\;\phi(\nu)}\; \Phi(r)\; r\; d\theta\; dr 
\;.
\end{displaymath}

For a distant observer, to whom the photodisk will appear as an
unresolved source, the total line intensity profile $I(\nu)$ is
obtained integrating all the infinitesimal contributions over the
photodisk:

\begin{displaymath}
I(\nu) = \int_0^{r_{ej}} \int_0^{2\pi} 
 e^{-K\; N(r,\theta)\;\phi(\nu)}\; \Phi(r)\; r\; d\theta\; dr \;.
\end{displaymath}

Given the definition of equivalent width

\begin{displaymath}
EW = \int_0^{+\infty} \left( 1-\frac{I(\nu)}{I_0}\right ) \; d\nu \;,
\end{displaymath}

\noindent 
the composite equivalent width can be expressed as follows:

\begin{displaymath}
EW= \int_0^{+\infty}\left ( 1 -   \int_0^{r_{ej}} \int_0^{2\pi} 
 e^{-K\; N(r,\theta)\;\phi(\nu)}\; \Phi(r)\; r\; d\theta\; dr \right) \; d\nu \;.
\end{displaymath}

Because of the normalization of $\Phi(r)$, the previous relation can
be rewritten as

\begin{equation}
\label{eq:ewfull}
EW=\int_0^{r_{ej}}\int_0^{2\pi}\int_0^{+\infty}
\left (1-e^{-K\; N(r,\theta)\;\phi(\nu)}\right )\;
\Phi(r,\theta)\;r\;d\theta\;dr\; d\nu\;
\end{equation}

Now, the inner integral is the equivalent width one would observe if
the physical system were composed only by the infinitesimal cloud
element $dA$ and the corresponding portion of the photodisk:

\begin{displaymath}
EW(r,\theta) = \int_0^{+\infty} 
\left (1-e^{-K\; N(r,\theta)\;\phi(\nu)}\right )\;d\nu\;,
\end{displaymath}

\noindent which implies that the composite equivalent width is the
weighted sum of the equivalent widths produced within each
infinitesimal cloud element.  If $g(N,b)$ is the curve of growth for
the given transition (where $b$ is the Doppler parameter that
characterizes the line profile), then Equation~\ref{eq:ewfull} can be
reformulated as:

\begin{equation}
\label{eq:ewsum}
EW=\int_0^{r_{ej}}\int_0^{2\pi}
g[N(r,\theta),b(r,\theta)] \; \Phi(r,\theta)\;r\;d\theta\;dr \;.
\end{equation}

%%%%%%%%%%%%%%%%%%%%%%%%%%%%%%%%%%%%%%%%%%%%%%%%%%%%%%%%%%%%%%%%%%%%%
%%%%%%%%%%%%%%%%%%%%%%%%%%%%%%%%%%%%%%%%%%%%%%%%%%%%%%%%%%%%%%%%%%%%%

\begin{acknowledgements}
The authors wish to thank L. Tacconi-Garman, S. Stanimirovi\'c and
A. Deshpande for their kind help. The authors are also grateful to
an anonymous referee for the useful comments and suggestions.
\end{acknowledgements}

\end{document}